\documentclass[aps,prd,showpacs,preprintnumbers,nofootinbib,twocolumn]{revtex4}
\usepackage[dvips]{graphicx}
\usepackage{bm,latexsym,amsmath,amssymb,amsfonts}
\begin{document}
\title{Can higher curvature corrections cure the singularity problem in $f(R)$ gravity?}

\author{Tsutomu~Kobayashi$^{1}$}
\email[Email: ]{tsutomu"at"gravity.phys.waseda.ac.jp}
\author{Kei-ichi~Maeda$^{1\,,2}$}
\email[Email: ]{maeda"at"waseda.jp}
\address{\,\\ \,\\
$^{1}$ Department of Physics, Waseda University,
Okubo 3-4-1, Shinjuku, Tokyo 169-8555, Japan\\
$^{2}$ Advanced Research Institute for Science and Engineering,
 Waseda University,
Okubo 3-4-1, Shinjuku, Tokyo 169-8555, Japan}

\begin{abstract}
Although $f(R)$ modified gravity models can be made to satisfy solar system and
cosmological constraints,
it has been shown that they have the serious drawback
of the nonexistence of stars with strong gravitational fields.
In this paper,
we discuss whether or not higher curvature corrections can remedy 
the nonexistence 
consistently.
The following problems are shown to arise as the
costs one must pay for the $f(R)$ models that allow for neutrons stars:
(i) the leading correction must be fine-tuned to have the typical energy scale
$\mu \lesssim 10^{-19}$ GeV, which essentially comes from the free fall time of a relativistic star;
(ii)
the leading correction must be further fine-tuned so that
it is not given by the quadratic curvature term.
The second problem is caused because
there appears an intermediate curvature scale,
and laboratory experiments of gravity will be under the influence of
higher curvature corrections.
Our analysis thus implies that it is a challenge to construct viable 
$f(R)$ models
without very careful and unnatural fine-tuning.
\end{abstract}

\preprint{WU-AP/295/08}
\maketitle

\section{Introduction}

The origin of the current accelerated expansion of the Universe~\cite{Acceleration} is
one of the biggest mystery in cosmology.
The accelerated expansion may be driven by some unknown energy-momentum component.
A more intriguing possibility is that
the acceleration could be due to long distance modification of gravity.
A simple class of modified gravity theories can be constructed
by generalizing the Einstein-Hilbert Lagrangian to some function
of the Ricci scalar, $f(R)$~\cite{Rev}.
Various models of $f(R)$ gravity have been proposed~\cite{CDTT, NO},
but inappropriate choices of the function readily
cause unwanted instability~\cite{DK} or gross violation of solar system
constraints~\cite{Chiba, ESK, CSE}.
The troubles arise due to an extra propagating scalar degree of freedom,
and hence viable $f(R)$ models must be constructed
in such a way that the dynamics of this scalar field is
carefully controlled. 
This is in principle possible, and indeed
$f(R)$ theories can be made to
satisfy solar system and laboratory tests by invoking
the chameleon mechanism~\cite{Chameleon, Cham2, Cham3}.
The key ingredient of the chameleon mechanism is the
density-dependent mass of the scalar field;
it mediates a short-range force in
high density environments such as the solar interior and vicinity.
(The actual mechanism to hide the chameleon field is slightly more involved~\cite{Chameleon}.)
Concrete examples of ``chameleon $f(R)$'' are found
in~\cite{Tegmark, St, Hu, AB} (see also Refs.~\cite{Cembranos:2005fi, Hall, NvA, Li, AT}).
They are the only known examples of viable $f(R)$ models
that exhibit no problems and no pathologies in the weak gravity
regime~\cite{Tsujikawa1, Tsujikawa2, Brax}.\footnote{
The model of~\cite{Tegmark}, which belongs to a different class of the models~\cite{St, Hu, AB},
is hardly distinguishable from $\Lambda$CDM cosmology
because of the very strong experimental constraints~\cite{Tsujikawa2}.
}

However, the potentially viable models of~\cite{St, Hu, AB} turn out to have
a serious drawback in the strong gravity regime.
That is, a deep (but not diverging) gravitational potential
drives the effective scalar degree of freedom to a curvature singularity.
This problem was first pointed out by Appleby and Battye in a cosmological setting~\cite{ABCos}
and then discussed by Frolov in a general context~\cite{Frolov}.
In the previous paper~\cite{KM}, we have studied
relativistic stars in $f(R)$ gravity and shown explicitly that
stars with strong gravitational fields develop curvature singularities and hence are prohibited.
The critical value of the potential is typically given by $|\Phi|\sim0.1$,
implying problematic nonexistence of neutron stars in the models of~\cite{St, Hu, AB}.

In this paper, we continue our program of studying strong gravity aspects 
of $f(R)$ gravity, and discuss whether or not higher curvature corrections to the original models
can resolve the singularity problem.
This is done again by constructing relativistic star solutions.
A higher curvature correction changes
the structure of the effective potential for the scalar degree of freedom
around the singularity~\cite{St, Dev}.
We consider a modified version of Starobinsky's $f(R)$~\cite{St},
adding a correction term proportional to $R^m$ $(m\ge2)$.
We also check whether or not
the chameleon mechanism works to pass local gravitational tests
in this modified $f(R)$ model.
Although we focus on the specific model,
our result will hold in the other similar models of this class.

This paper is organized as follows:
In the next section, we describe the field equations of $f(R)$ modified gravity
in terms of a scalar-tensor theory.
Then, in Sec.~\ref{sec:high}, we define the specific theory we consider.
Our numerical results are presented in Sec.~\ref{sec:num}.
In Sec.~\ref{sec:fine}, we
discuss local tests of gravity in the $f(R)$ model
and point out the problem associated with
the high energy correction term.
We draw our conclusions in Sec.~\ref{sec:conc}.

\section{$f(R)$ gravity as a scalar-tensor theory}

\subsection{Field equations}

The action we consider has the form of
\begin{eqnarray}
S=\int d^4x\sqrt{-g}\left[\frac{f(R)}{16\pi G}+{\cal L}_{\rm m}\right],
\label{action}
\end{eqnarray}
where
$f(R)$ is a function of the Ricci scalar $R$, and
${\cal L}_{{\rm m}}$ is the Lagrangian of matter fields.
Variation with respect to metric leads to the field equations\footnote{In this paper, 
we focus on the metric approach rather than the Palatini one.}
\begin{eqnarray}
f_R R_{\mu\nu}-\nabla_{\mu}\nabla_{\nu}f_R
+\left(\Box f_R-\frac{1}{2}f\right)g_{\mu\nu}=8\pi G T_{\mu\nu},
\label{field_eq_original}
\end{eqnarray}
where $f_R:=df/dR$ and $T_{\mu\nu}:=-2\delta{\cal L}_{\rm m}/\delta 
g^{\mu\nu}+g_{\mu\nu}{\cal L}_{\rm m}$.
The trace of  Eq.~(\ref{field_eq_original}) reduces to
\begin{eqnarray}
\Box f_R=\frac{8\pi G}{3} T+\frac{1}{3}(2f-f_RR).\label{feqtr}
\end{eqnarray}
We now introduce an effective scalar degree of freedom, which sometimes is dubbed ``scalaron,'' by
defining $\chi:=f_R$.
Inverting this relation, the Ricci scalar can be expressed in terms of
$\chi$: $R=Q(\chi)$.
In this way Eqs.~(\ref{field_eq_original}) and~(\ref{feqtr}) are equivalently 
rewritten as~\cite{BD}
\begin{eqnarray}
\chi G_{\mu}^{\;\nu}
&=&8\pi G T_{\mu}^{\;\nu}+\left(\nabla_{\mu}\nabla^{\nu}
-\delta_{\mu}^{\;\nu}\Box\right)\chi
-\chi^2 V(\chi)\delta_{\mu}^{\;\nu},
\label{fieldeq}
\\
\Box\chi&=&\frac{8\pi G}{3}T+\frac{2\chi^3}{3}\frac{dV}{d\chi},
\label{scalareq}
\end{eqnarray}
where the potential $V$ is given by
\begin{eqnarray}
V(\chi):=\frac{1}{2\chi^2}\left[\chi Q(\chi)-f(Q(\chi))\right],
\end{eqnarray}
and $dV/d\chi =\left[2f(Q(\chi))-\chi Q(\chi)\right]/(2\chi^3) $.

Eqs.~(\ref{fieldeq}) and~(\ref{scalareq}) are equivalent to
the Jordan frame equations of motion
in the  Brans-Dicke theory with $\omega=0$
plus a potential $V(\chi)$.
One can move to the Einstein frame
by performing the conformal transformation
$\tilde g_{\mu\nu} = \chi g_{\mu\nu}$ with $\chi = \exp (\sqrt{16\pi G/3}\,\phi )$,
where $\phi$ is the canonical scalar field.
The potential for $\phi$ is then given by $V(\chi(\phi))$.
However, we do not work in the Einstein frame in the following discussion.

\subsection{Classical mechanical analogy}

We are going to investigate static, spherically symmetric stellar solutions in the above system.
To study the radial profile of $\chi$ through Eq.~(\ref{scalareq}),
it is useful to note that the equation can be written as
\begin{eqnarray}
\frac{d^2\chi}{dr^2}+\frac{2}{r}\frac{d\chi}{dr}=-\frac{dU}{d\chi}+{\cal F},
\label{cmanalog}
\end{eqnarray}
where
\begin{eqnarray}
\frac{dU}{d\chi}=-\frac{2\chi^3}{3}\frac{dV}{d\chi}
\end{eqnarray}
and
${\cal F} =(8\pi G/3)  T$.
Here we have ignored the effect of the metric for simplicity.
(Later we will solve the full set of the field equations numerically.)
Now, by identifying $r$ as a time coordinate,
Eq.~(\ref{cmanalog}) can be regarded as
the equation of motion in classical mechanics.
One can understand the radial profile of $\chi$ intuitively
as the motion of a particle in the potential $U$ under the time-dependent force ${\cal F}$
(and the frictional force corresponding to the second term in the left hand side).
The mechanical analogy is particularly useful to
comprehend the essential point  of
the nonexistence statement for relativistic stars
in $f(R)$ gravity~\cite{KM}.


\section{Adding higher curvature corrections to $f(R)$ gravity}\label{sec:high}

\begin{figure}[tb]
  \begin{center}
    \includegraphics[keepaspectratio=true,height=50mm]{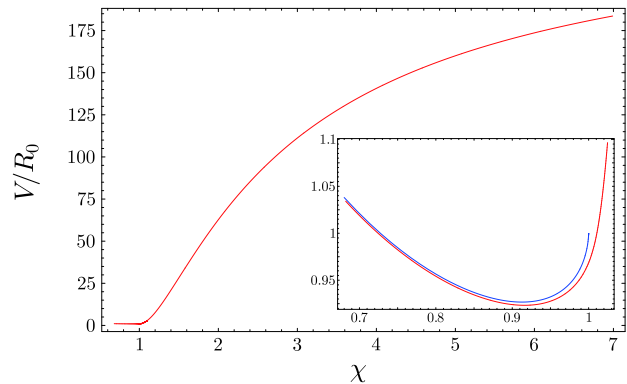}
  \end{center}
  \caption{The potential $V$. The inset shows the structure around the de-Sitter minimum.
  The potential of the original model (without $R^m$ term) is shown by a blue line
  for purpose of comparison.
  Parameters are
  given by $\lambda=2$, $n=1$, $m=2$, and $\varepsilon=5\times10^{-4}$.
  The point $\chi=1$ corresponds to a curvature singularity in the original model,
  but the $R^m$ term pushes the curvature singularity toward infinity, $\chi=\infty$.}%
  \label{fig:vm2.eps}
\end{figure}
\begin{figure}[tb]
  \begin{center}
    \includegraphics[keepaspectratio=true,height=52mm]{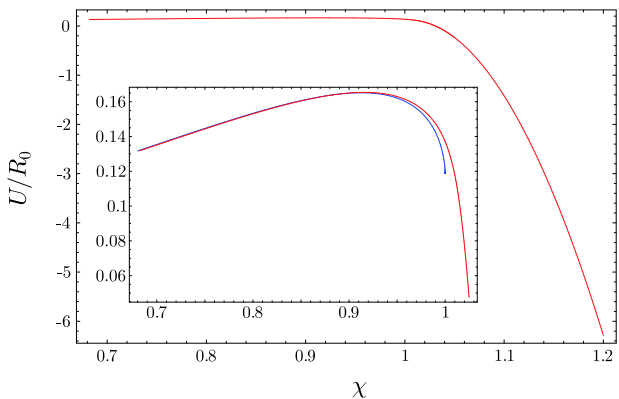}
  \end{center}
  \caption{The effective potential $U$.
  The inset shows the structure around the de-Sitter extremum.
  The effective potential of the original model (without $R^m$ term) is shown by a blue line
  for purpose of comparison.  Parameters are
  given by $\lambda=2$, $n=1$, $m=2$, and $\varepsilon=5\times10^{-4}$.
  The dangerous curvature singularity is pushed toward $\chi=\infty$
  by the $R^m$ term.}
  \label{fig:um2.eps}
\end{figure}

In the previous paper~\cite{KM} we studied
the strong gravity aspect of Starobinsky's $f(R)$ theory described by
$f(R)=R+\lambda R_0[(1+R^2/R_0^2)^{-n}-1]$~\cite{St}.
There we showed that
stars with strong gravitational fields (e.g., neutron stars) cannot exist
in this model.
We argued that 
this statement applies to the other similar models~\cite{Hu, AB} as well.
This problem arises due to the dynamics of the effective scalar degree of freedom, $\chi$,
in the high curvature regime.
Therefore, the problem may be cured by adding higher curvature corrections
that modify the structure of the potential around the large $R$ region,
as already noted in the original reference~\cite{St}
and later discussed in~\cite{Dev}.

In general,
higher curvature corrections may be written as
$a_2R^2+a_3R^3+\cdots$, and so the most natural choice of
the leading order term will be $R^2/\mu^2$.\footnote{
Higher order corrections naturally include terms like $R_{\mu\nu}R^{\mu\nu}$,
but in this paper we focus on the $f(R)$-type modified gravity and hence simply
assume that the corrections are also given by a function of the Ricci scalar.}
The $R^2$ term may be responsible for inflation in the {\em early} Universe
if $\mu$ is set to be an inflationary scale (e.g., $\mu\sim10^{12}\,$GeV)~\cite{StInf},
but in this paper we do not restrict the mass scales 
of the curvature correction and
assume that such parameters in the high energy correction terms 
take rather arbitrary values.
If the coefficient of the $R^2$ term is highly suppressed for some reason, then
the leading correction will be the form of $R^3/\mu^4$.\footnote{Note that 
the higher curvature correction is given by the quartic 
terms in type II superstring theory.}
To make the model simple but general enough,
let us consider a function
\begin{eqnarray}
f(R)=R+\lambda R_0\left[\left(1+\frac{R^2}{R_0^2}\right)^{-n}-1\right]+\frac{R^m}{\mu^{2(m-1)}},
\label{fRSt}
\end{eqnarray}
where $n \,(>0), \lambda \,(>0)$, $R_0 \, (>0)$, and $m\,(\geq 2)$ are parameters.
The present Hubble scale is basically given by $H_0^2\sim{\cal O}(R_0)$.
We define a dimensionless parameter $\varepsilon:= R_0/\mu^2$ and
assume that $\varepsilon \ll 1$ since
the last term in Eq.(\ref{fRSt}) is the high energy correction.
At sufficiently low energies we
have no cosmological constant, $f(R)\simeq R$, while for $R_0\ll R\ll \mu^2$ we find
$f(R)\simeq R-\lambda R_0 +\lambda R_0^{2n+1}/R^{2n}+\cdots$.
At very high energies, $R\gg\mu^2$, the last term dominates.

A de Sitter solution, $R=R_1=$ constant, minimizes the potential $V(\chi)$, and hence
is found by solving the algebraic equation
\begin{eqnarray}
2f(R_1)-R_1f_{R}|_{R=R_1} = 0.
\end{eqnarray}
We may define the effective ``cosmological constant'' as $\Lambda_{{\rm eff}}:=R_1/4$.

The scalar field $\chi$ is written in terms of $R$ as
\begin{eqnarray}
\chi=1-2n\lambda\frac{R}{R_0}\left(1+\frac{R^2}{R_0^2}\right)^{-n-1}+m\left(\frac{R}{\mu^2}\right)^{m-1}.
\label{chi=R+}
\end{eqnarray}
In the original model without the $R^m$ correction,
a curvature singularity $R=\infty$ corresponds
to a finite $\chi$ ($\chi=1$) and this is very close to the de Sitter minimum, $\chi_1 = \chi(R_1)$.
However, as is clear from
Eq.~(\ref{chi=R+}), the dangerous curvature singularity now corresponds to $\chi=\infty$,
and hence one may expect that this model is safe.
A typical form of the potential $V(\chi)$ is shown in Fig.~\ref{fig:vm2.eps}.
The effective potential $U(\chi)$ is also plotted in Fig.~\ref{fig:um2.eps}.
A straightforward calculation shows
$V\propto R^{-m+2}$ for $R\gg\mu^2$.
Therefore, $V\to$ const. as $R\to\infty$ for $m=2$,
while $V\to0$ in the same limit for $m\ge3$.
Similarly, we have
\begin{eqnarray}
3\frac{dU}{d\chi}\approx-R+(m-2)\frac{R^m}{\mu^{2(m-1)}}
\quad (R\gg R_0).
\end{eqnarray}
From this we see that in the $m=2$ case
the effective potential $U$ becomes steeper as the curvature increases,
leading finally to $dU/d\chi\to-\infty$ as $R\to+\infty$.
For $m\ge3$, $U$ has a minimum at $R\sim\mu^2$ and $dU/d\chi\to+\infty$ as $R\to+\infty$.



\section{Relativistic stars in $f(R)$ gravity with high energy corrections}\label{sec:num}

\begin{figure}[tb]
  \begin{center}
    \includegraphics[keepaspectratio=true,height=55mm]{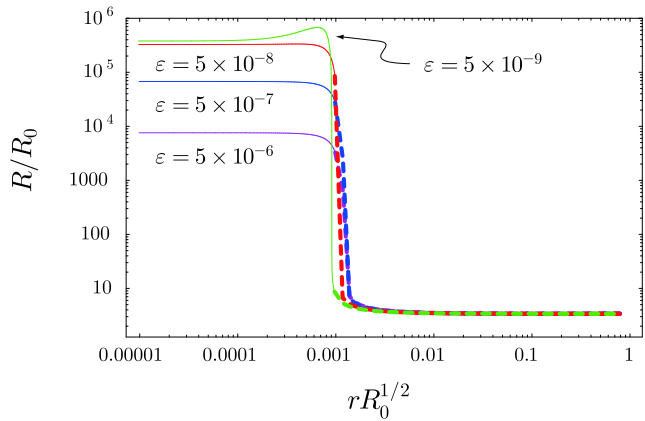}
  \end{center}
  \caption{Plots of the Ricci scalar $R(r)$ for different $\varepsilon$. Parameters are given by
  $\lambda=2$, $n=1$, $m=2$. The energy density is $4\pi G\rho_0=10^6\Lambda_{{\rm eff}}$
  and the central pressure is $p_c = 0.3\rho_0$.
  Solid (dashed) lines correspond to the region inside (outside) the star.
  These examples typically give $\hat G M/{\cal R} \simeq 0.25$ -- $0.26$.}%
  \label{fig:curvature.eps}
\end{figure}

\begin{figure}[tb]
  \begin{center}
    \includegraphics[keepaspectratio=true,height=55mm]{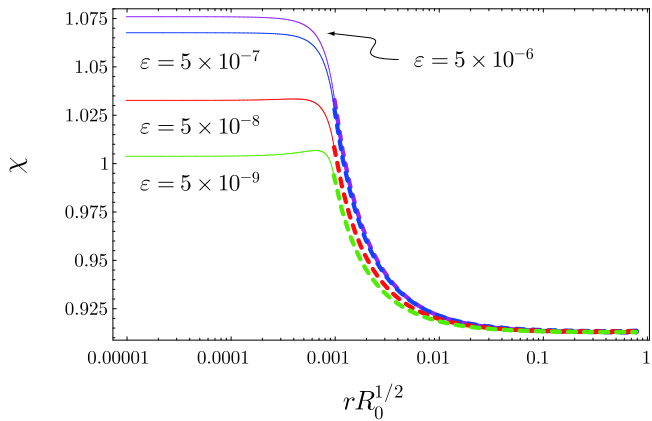}
  \end{center}
  \caption{Plots of $\chi(r)$ for different $\varepsilon$. Parameters are the same as
  those in Fig.~\ref{fig:curvature.eps}.
  Solid (dashed) lines correspond to the region inside (outside) the star.}%
  \label{fig:chi.eps}
\end{figure}

\begin{figure}[tb]
  \begin{center}
    \includegraphics[keepaspectratio=true,height=55mm]{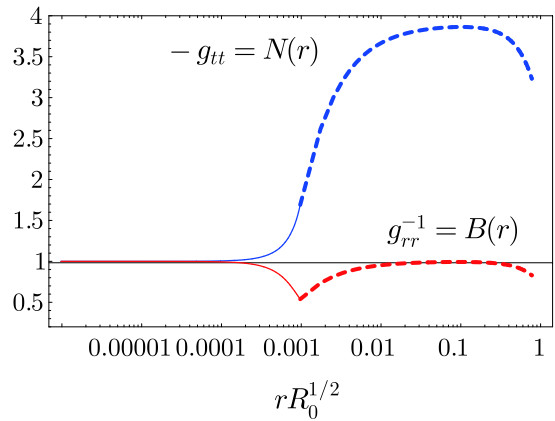}
  \end{center}
  \caption{Plots of the metric functions for $\varepsilon=5\times10^{-9}$.
  Solid (dashed) lines correspond to the region inside (outside) the star.}%
  \label{fig:metric.eps}
\end{figure}

\begin{figure}[tb]
  \begin{center}
    \includegraphics[keepaspectratio=true,height=55mm]{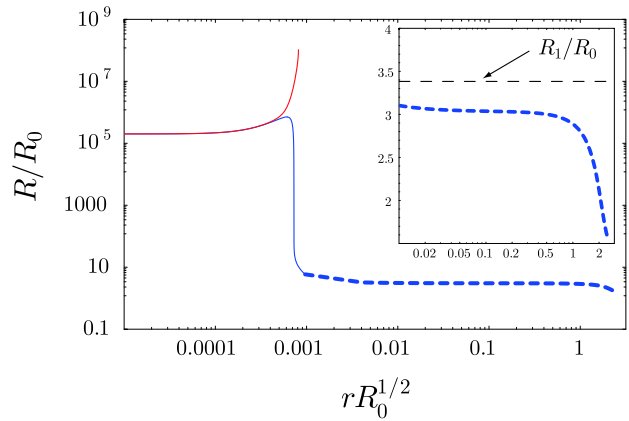}
  \end{center}
  \caption{Plots of $R(r)$ for $\varepsilon=5\times10^{-10}$. Parameters are the same as
  those in Fig.~\ref{fig:curvature.eps}.
  The upper (red) line is a plot for the solution
  with the central curvature $R_c = 0.1185\times 8\pi G\rho_0$.
  This solution exhibits the behavior of falling toward the singularity.
  The lower (blue) line indicates a ``overshooting'' solution with $R_c = 0.1184\times 8\pi G\rho_0$.
  The inset emphasizes the overshooting behavior. 
  }%
  \label{fig:rolling.eps}
\end{figure}

We now investigate static and spherically symmetric stars with constant densities
(i.e., a generalization of the Schwarzschild interior solution) in the model
defined by Eq.~(\ref{fRSt}).
We shall work along the lines of the previous paper~\cite{KM}.
The basic equations are found there and are replicated in Appendix~\ref{app:be}.
Stars in $f(R)$ gravity have been studied also in Ref.~\cite{stars}.

Given a density $\rho_0$ and the central values of the pressure $p_c$ and
the scalar field $\chi_c$ (or, equivalently, the central curvature $R_c$),
we can integrate Eqs.~(\ref{dp=})--(\ref{eq_chi}) numerically from the regular center $r=0$
to the surface of the star, $r={\cal R}$, defined by $p({\cal R})=0$.
(The boundary condition at the center is also given
in Appendix~\ref{app:be}.)
Then, imposing the continuity of the metric functions $N(r)$ and $B(r)$,
the scalar field $\chi$, and its derivative $d\chi/dr$ at the stellar surface,
we integrate the vacuum field equations~(\ref{fe_B})--(\ref{eq_chi})
to find the exterior geometry.
We are looking for a solution such that it is asymptotic to de Sitter with $\Lambda_{{\rm eff}}=R_1/4$
(and hence $\chi\to\chi_1$).
For fixed $\rho_0$ and $p_c$, we can find
the desired solution (if it exists) by carefully tuning the initial value $\chi_c=\chi_{{\rm crit}}$.
In the mechanical analogy, this solution corresponds to the situation where
the particle starts at rest and reaches the top of the potential $(\chi=\chi_1)$ in the limit of $r\to\infty$.
The particle overshoots the top of the potential for $\chi_c<\chi_{{\rm crit}}$,
while it turns around before it reaches the top and falls into the singularity
 for $\chi_c>\chi_{{\rm crit}}$.

In the previous paper~\cite{KM} we showed
that $\chi_{{\rm crit}}$ becomes larger as the gravitational potential of the star increases,
getting eventually at $\chi_s$, above which the slope of the potential $dU/d\chi$ is
greater than the force term and so the particle cannot climb up the potential.
Thus, if the gravitational potential is larger than a certain value,
the desired solution described above does not exist,
and we only have two types of singular solutions, i.e., a ``falling-down"
type and a ``overshooting'' one.

Bearing the above mechanical picture in mind,
let us move on to the case with the $R^m$ correction.
We have carried out numerical calculations
for various values of $\varepsilon=R_0/\mu^2$ with
fixing the other model parameters as $\lambda=2$, $n=1$, and $m=2$. 
The profiles of $\chi$ and $R$ for regular, asymptotically de Sitter solutions
are shown in Figs.~\ref{fig:curvature.eps} and~\ref{fig:chi.eps},
and the metric functions for the $\varepsilon=5\times 10^{-9}$ case
are plotted in Fig.~\ref{fig:metric.eps}.
In these plots
the energy density and the central pressure are given respectively by
$4\pi G\rho_0=10^6\Lambda_{{\rm eff}}$ and $p_c = 0.3\rho_0$,
leading to the gravitational potential as large as $\hat G M/{\cal R} 
\sim 0.25$,
where $\hat G:=G/\chi_c$ and $M:=4\pi\rho_0{\cal R}^3/3$.
(The gravitational potential is controlled by the ratio $p_c/\rho_0$.)
Stars with such large potentials are prohibited in the original model without
the high energy correction.

\begin{figure}[tb]
  \begin{center}
    \includegraphics[keepaspectratio=true,height=74mm]{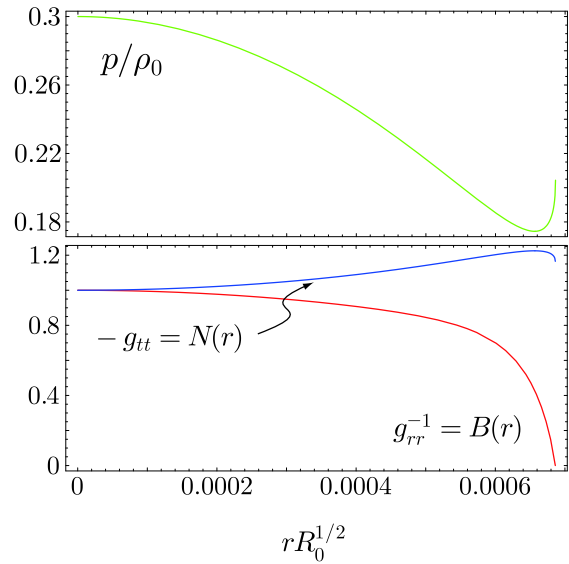}
  \end{center}
  \caption{If the force ${\cal F}$ is too weak to defeat the potential slope $dU/d\chi$
  inside a star and consequently
  $\chi(r)$ and $R(r)$ grow without turning back, the pressure shows an unphysical behavior.
  The coordinate choice is not good in this case, as can be seen most clearly from
  the behavior of the metric component $g_{rr}$.
  Plots are for $\lambda=2$, $n=1$, $4\pi G\rho_0=10^6\Lambda_{{\rm eff}}$, 
  $p_c=0.3\rho_0$, and
  $R_c=0.12\times8\pi G\rho_0$.}%
  \label{fig:singpr.eps}
\end{figure}

For $\varepsilon < 5\times10^{-9}$, however, 
we find that the regular solution ceases to exist.
To see this, we show
the behavior of the Ricci scalar for $\varepsilon=5\times10^{-10}$ in 
Fig.~\ref{fig:rolling.eps}.
Taking $R_c = 0.1185\times 8\pi G\rho_0\, [=1.185\times8\pi G(\rho_0-3p_c)]$,
$\chi$ goes toward $\chi=\infty\,(R=\infty)$.
In this case, the pressure
is an increasing function of $r$ away from the center, and
there is a maximal circumferential radius corresponding
to an infinite proper distance from the center, as can be seen from 
Fig.~\ref{fig:singpr.eps}.
It is not an asymptotically flat space, but a ``cylindrical" shape
 space with a singularity at the end.
Taking a slightly smaller value, $R_c = 0.1184\times 8\pi G\rho_0$,
$\chi$ then overshoots the top of the potential and rolls down to the left.
During this rolling-down phase the Kretschmann scalar, 
$R_{\mu\nu\rho\sigma}R^{\mu\nu\rho\sigma}$,
diverges.
Only these two cases are realized,
both of which are unphysical.
This is the same situation one encounters in the model without the high energy correction~\cite{KM}.
In order for the higher curvature term to come to the rescue,
it must be sufficiently large.

\begin{figure}[tb]
  \begin{center}
    \includegraphics[keepaspectratio=true,height=55mm]{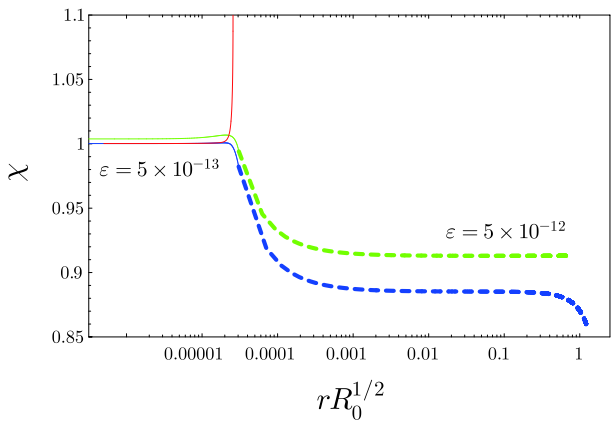}
  \end{center}
  \caption{Plots of $\chi(r)$ for $4\pi G\rho_0=10^9\Lambda_{{\rm eff}}$ and $p_c=0.3\rho_0$.
  The model parameters are given by $\lambda=2$, $n=1$, and $m=2$.
  In the $\varepsilon=5\times10^{-12}$ case, the desired solution is obtained for
  $R_c=0.2245\times8\pi G\rho_0$.
  However, in the case of $\varepsilon=5\times10^{-13}$ only singular solutions are
  found. Two examples are shown: one is for $R_c=0.1185\times8\pi G\rho_0$
  (falling rapidly down to $R=\infty$)
  and the other is for $R_c=0.1184\times8\pi G\rho_0$ (a overshooting solution).}%
  \label{fig:hd.eps}
\end{figure}
\begin{figure}[tb]
  \begin{center}
    \includegraphics[keepaspectratio=true,height=70mm]{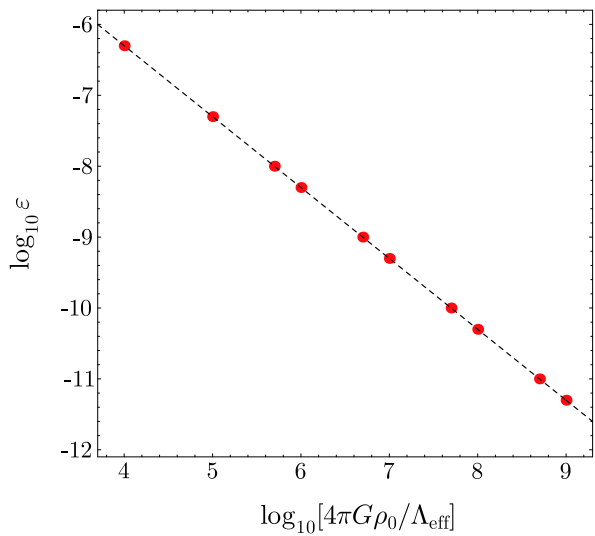}
  \end{center}
  \caption{The minimum value of $\varepsilon$ as a function of $\rho_0$.
  Points indicate numerical results, showing the scaling relation.}%
  \label{fig:scaling.eps}
\end{figure}

The minimum value of $\varepsilon$ (or the maximum value of $\mu^2$)
that allows for relativistic stars depends on the energy density.
To explore the bound on $\mu^2$,
we have performed numerical calculations for different values of $\rho_0$ 
ranging
from $\rho_0=10^4\Lambda_{{\rm eff}}/(4\pi G)$ to
$\rho_0=10^9\Lambda_{{\rm eff}}/(4\pi G)$.\footnote{To mimic a neutron star,
the density must be $8\pi G\rho_0\sim 10^{44}\Lambda_{{\rm eff}}$.
However, it is difficult to implement such an extremely high density contrast
in our numerical computations.}
As an example,
the behavior of $\chi$ for $\rho_0=10^9\Lambda_{{\rm eff}}/(4\pi G)$ is shown in Fig.~\ref{fig:hd.eps}.
From our numerical results it is
confirmed that the minimum value of $\varepsilon$ is inversely proportional to the energy density
(Fig.~\ref{fig:scaling.eps}),
and we roughly have the bound
$\varepsilon\gtrsim10^{-2}R_0/(8\pi G\rho_0)$
in order for stars with strong gravitational fields ($\sim0.25$) 
to exist.\footnote{For given
$\varepsilon$ there is a maximal gravitational potential.
Therefore, the minimum value of $\varepsilon$ will be different depending on
how large gravitational potentials one needs.}
This condition gives
\begin{eqnarray}
\mu^2< \alpha\times 8\pi G \rho_0,\quad \alpha\sim{\cal O}(10^2).
\label{result}
\end{eqnarray}
Taking $\rho_0\sim\rho_{{\rm nucl}}\sim 10^{14}\,$g/cm$^3\sim 10^{-3}\,$GeV$^4$
(nuclear density),
one arrives at
$\mu\lesssim10^{-19}\,$GeV.
This result itself is not surprising because $8\pi G\rho_{{\rm nucl}}$ is a natural scale
associated with neutron stars.
However, purely from a theoretical point of view,
this provides an {\em unnaturally small} energy scale.
Obviously, the $R^2/\mu^2$ term with such small $\mu$
cannot be relevant to inflation in the early Universe.

We have also done numerical calculations to construct stellar solutions
with $\hat GM/{\cal R}\sim 0.26$ in the $m=3$ model,
and obtained
essentially the same result:
for sufficiently large $\varepsilon$ we can find regular, asymptotically de Sitter solutions,
while for $\varepsilon$ smaller than a certain value we only have
two classes of unphysical solutions.
Note here that for $m\ge 3$ the structure of the (effective) potential near the curvature singularity is
quite different from that of the $m=2$ model.
Nevertheless, the solution corresponding to $\chi(r)$ moving toward right
shows an unphysical nature: the pressure is an increasing function of $r$ away from the central region
and $g_{rr}\to\infty$ at finite $r$.
To allow for relativistic stars with $\hat GM/{\cal R}\sim 0.26$, it is required that
$\mu^2< \alpha'\times 8\pi G \rho_0$ where $\alpha'\sim{\cal O}(10)$.

Before closing this section,
let us comment on the behavior of the metric for asymptotically de Sitter stellar solutions.
A numerical fitting leads to the approximate expression for the metric outside stars:
\begin{eqnarray}
N&\simeq&N_{\infty}\left(1-2c_1\frac{{\cal R}}{r}-\frac{c_2}{3}\Lambda_{{\rm eff}}r^2\right),
\\
B&\simeq& 1-2c_3\frac{{\cal R}}{r}-\frac{c_4}{3}\Lambda_{{\rm eff}}r^2,
\end{eqnarray}
where $c_2\simeq c_4\simeq 1.0$ irrespective of $\varepsilon$ $(\gtrsim5\times 10^{-9})$,
while $c_1$ and $c_3$ are slightly different for different $\varepsilon$.
For $\varepsilon=5\times10^{-9}$ one finds
$c_1\simeq0.29$ and $c_3\simeq0.24$. This gives the post-Newtonian
parameter $\gamma\simeq c_3/c_1\simeq0.81$.
For $\varepsilon=5\times10^{-6}$
one has $c_1\simeq0.30$, $c_3\simeq0.21$, and $\gamma\simeq0.72$. 
These results imply that the chameleon mechanism does not work
in the above examples.
This is because we are considering a vacuum exterior.
Note that taking into account the effect of surrounding media
does not remedy the nonexistence of relativistic stars
for small $\varepsilon$: to avoid falling down toward large $R$,
$\chi$ inevitably overshoots the top of the potential
also in the presence of exterior matter as we have ${\cal F}<0$ outside the star.

\section{Appearance of an intermediate scale and another fine-tuning}\label{sec:fine}

Assuming that the ``UV scale'' is given by $\mu^2\sim8\pi G\rho_{{\rm nucl}}$,
one may naively expect that higher curvature corrections have no impact on
local tests of gravity since relevant densities are
much smaller.
However, as we show below, this expectation is not true.
The purpose of this section is to point out a new problem
brought by the higher curvature correction.

To discuss the behavior of gravity in laboratories,
it is convenient to define the effective potential for the $\chi$ field as
\begin{eqnarray}
\frac{dV_{{\rm eff}}}{d\chi}=\frac{1}{3}\left[
2f(Q(\chi))-\chi Q(\chi)
\right]+\frac{8\pi G}{3}T,
\end{eqnarray}
where
the energy-momentum tensor of matter is included.
We consider the regime $R_0\ll R\ll\mu^2$.
In this regime Eq.~(\ref{chi=R+}) can be written as
\begin{eqnarray}
\chi \approx 1-2n\lambda\left(\frac{R_0}{R}\right)^{2n+1}
+m\left(\frac{R}{\mu^2}\right)^{m-1}.
\end{eqnarray}
At the minimum of the effective potential $V_{{\rm eff}}$
one finds
\begin{eqnarray}
\frac{dV_{{\rm eff}}}{d\chi}=0
\;
\Rightarrow \;
R\approx 8\pi G\rho,
\end{eqnarray}
where $\rho\approx-T$ is the energy density of nonrelativistic matter.
The mass of the excitation of the $\chi$ field around the minimum is given by
$m_{\chi}^2 =d^2V_{{\rm eff}}/d\chi^2|_{R\approx8\pi G\rho}$.
Thus, the Compton wavelength $\lambda_\chi =m_\chi^{-1}$ can be computed as
\begin{eqnarray}
\lambda_{\chi}^2&\approx &3f_{RR}|_{R\approx8\pi G\rho}
\nonumber\\
&\approx& \left.\frac{k_1}{R}\left(\frac{R_0}{R}\right)^{2n+1} 
+\frac{k_2}{R}\left(\frac{R}{\mu^2}\right)^{m-1}\right|_{R\approx8\pi G\rho},
\label{compt}
\end{eqnarray}
where
$k_1:=6n(2n+1)\lambda$
and
$k_2:=3m(m-1)$.

To evaluate the Compton wavelength,
it is important to note that
there is a critical curvature scale defined by
\begin{eqnarray}
R_*:=\left(R_0^{2n+1}\mu^{2(m-1)}\right)^{1/(2n+m)},\label{intcur}
\end{eqnarray}
and for $R\ll R_*$ (respectively, $R\gg R_*$) the first (respectively, second) term in Eq.~(\ref{compt})
is much greater than the other.
Eq.~(\ref{intcur})
gives an intermediate curvature scale $R_0\ll R_*\ll \mu^2$,
which implies that {\em the high energy correction term 
comes into play in determining
the Compton wavelength of $\chi$
at a much lower scale than expected}.
In terms of energy densities, the intermediate scale $\rho_*:=R_*/(8\pi G)$
may be written as
\begin{eqnarray}
\rho_*:=\left(\rho_{{\rm DE}}^{2n+1}\rho_{{\rm UV}}^{ m-1 }\right)^{1/(2n+m)},
\label{int_dens}
\end{eqnarray}
where $\rho_{{\rm DE}}\sim 10^{-30}\,$g/cm$^3$
and $\rho_{{\rm UV}}\lesssim \rho_{{\rm nucl}}\sim 10^{14}\,$g/cm$^{3}$.
For instance, putting $n=1$ and $m=2$
yields $\rho_*\lesssim 10^{-19}\,$g/cm$^3$.
Laboratory experiments are usually done at densities 
much higher than this!

Although the original model is made to satisfy
solar system and laboratory tests,
the intermediate scale brought by a high energy correction term
can destroy its success.
Indeed, in the above example ($n=1$ and $m=2$),
the mass of $\chi$ is independent
of local energy densities for $\rho\gg\rho_*$
and so the chameleon mechanism does not work in laboratories.
Since $\chi$ has a gravitational-strength coupling and
the Compton wavelength is evaluated as
$\lambda_\chi\sim \mu^{-1}\gtrsim10^5\,$cm,
the $m=2$ model is ruled out by the fifth force constraint~\cite{Will:2005va, Kapner:2006si}.

As seen from Eqs.~(\ref{compt}) and~(\ref{intcur}),
the intermediate scale and its consequences
are sensitive to the explicit form of the higher curvature correction.
For example, in the $n=1$ and $m=3$ case one obtains $\rho_*\lesssim10^{-12}\,$g/cm$^{3}$,
and for densities higher than this the Compton wavelength is found to be
$\lambda_\chi\gtrsim 0.1\,{\rm mm}\times(\rho/1\;{\rm g\cdot cm^{-3}})^{1/2}$.
This typically gives the marginal scale
tested by laboratory experiments of gravity.
Thus, determining whether or not a given high energy correction
satisfies local tests requires a more careful study, 
which is beyond the scope of the present paper. 
We just emphasize here that
``high energy corrections'' play a crucial role above the intermediate curvature scale, and
the $R^2/\mu^2$ correction, which seems to
appear in natural circumstances,
is clearly inconsistent with laboratory tests
if one chooses the parameter $\mu^2$ so that
the theory evades the nonexistence statement of neutron stars.

Finally,
let us comment on cosmology with the $R^m$ term.
In the matter-dominated era, we have an estimate $m^2_\chi/H^2\sim (\mu^2/H^2)^{m-1}\gg 1$,
where it is assumed that
the matter energy density is much greater than $\rho_*$.
This implies that the excitation of $\chi$ is suppressed,
rendering the field safe for cosmology.
Before the time of matter-radiation equality
the energy density of nonrelativistic matter is given by
$\rho_{{\rm m}}=\mathtt{r}\rho_{{\rm r}}$,
where $\rho_{{\rm r}}$ is the energy density of radiation and
$\mathtt{r}:=a/a_{{\rm eq}}\ll1$.
Since $R\sim 8\pi G\rho_{{\rm m}}$ and $H^2\sim 8\pi G\rho_{{\rm r}}/3$,
one ends up with
$m^2_{\chi}/H^2\sim\mathtt{r}^{-m+2}(\mu^2/H^2)^{m-1}$.
Even at nucleosynthesis, the ratio $m^2_{\chi}/H^2$
is enhanced by the factor $\mathtt{r}^{-m+2}$
except for $m=2$.
Thus, we can approximately reproduce standard cosmology for $H^2\lesssim\mu^2$
in models with $m\ge3$.

\section{Summary and Conclusions}\label{sec:conc}

In this paper, we have tried to resolve the singularity problem
arising in the strong gravity regime of otherwise viable $f(R)$ theories.
Adding a higher curvature correction in the form of $R^m/\mu^{2(m-1)}$,
we have studied stars with strong gravitational fields
which were prohibited in the original models.
Our numerical results have shown that
the scale $\mu^2$ cannot be taken to be
as large as an inflationary energy scale
nor a natural UV cutoff scale like $(8\pi G)^{-1}$.
Rather,
$\mu^2\lesssim{\cal O}(8\pi G\rho)$, where $\rho$ is
the stellar density and hence is taken to be a nuclear density, is required
in order to remedy the nonexistence of relativistic stars.
This provides a ``high''
energy scale as small as $\mu\lesssim 10^{-19}\,$GeV.
This is the first fine-tuning required for the high energy correction.

In contrast to the naive expectation,
the high energy corrections come into play
at an intermediate curvature scale in
determining the mass of $\chi$'s excitation around the minimum of
the effective potential.
If the leading correction is given by the quadratic curvature term,
the intermediate scale is
$
R_*\sim\left(R^{2n+1}_0\mu^2\right)^{1/(2n+2)}
$,
and the corresponding energy density is
$\rho_*\sim 10^{-19}\,$g/cm$^3$ for $n=1$.
For densities higher than this,
the Compton wavelength of $\chi$ is $\sim\mu^{-1}\sim 10^5\,$cm.
Therefore, the high energy correction
completely destroys the success of the original $f(R)$ model
that passes local tests of gravity.
If the quadratic correction is {\em suppressed}
relative to the other higher curvature terms,
possibly this is not always the case.
However, it might be thought of as a problem that gravity in 
the intermediate curvature regime ($R_0\ll R\ll \mu^2$)
is so sensitive to the explicit form of
UV correction terms.
In this sense we need another fine-tuning of the high energy correction.

To conclude, although
there is still a very small
room for a possible construction of viable $f(R)$ models
that evade local tests of gravity and allow for stars with strong gravitational fields, very careful and unnatural
fine-tuning is required
for the model construction, leaving
challenges for $f(R)$ modified gravity.

\acknowledgments

This work was partially supported by the JSPS under Contact No.~19-4199,
by the Grant-in-Aid for Scientific Research
Fund of the JSPS (No.~19540308) and by the
Japan-U.K. Research Cooperative Program.

\appendix

\section{Spherically symmetric stars in $f(R)$ gravity}\label{app:be}

In this appendix we summarize the basic equations
for constructing spherically symmetric stellar solutions in $f(R)$ gravity~\cite{KM}.

\subsection{Basic equations}

We take  the ansatz of a spherically symmetric and  static metric:
\begin{eqnarray}
ds^2=-N(r) dt^2+ \frac{dr^2}{B(r)}+r^2\left(d\theta^2+\sin^2\theta
 d\varphi^2\right).
\label{met}
\end{eqnarray}
The energy-momentum tensor of matter fields is given by
\begin{eqnarray}
T_{\mu}^{\;\nu}=\text{diag}\left(-\rho, p, p, p\right).
\end{eqnarray}
From
the energy-momentum conservation, $\nabla_{\nu}T_{\mu}^{\;\nu}=0$, we obtain
\begin{eqnarray}
p'+\frac{N'}{2N}(\rho+p)=0.\label{dp=}
\end{eqnarray}
Here and hereafter a prime denotes differentiation with respect to $r$.
The $(tt)$ and $(rr)$ components of the field equations~(\ref{fieldeq}) yield,
 respectively,
\begin{eqnarray}
&&\frac{\chi}{r^2}\left(-1+B+rB'\right)=-8\pi G\rho-\chi^2V
\nonumber\\&&\qquad\qquad\qquad
-B\left[\chi''+\left(\frac{2}{r}+\frac{B'}{2B}\right)\chi'\right],
\label{fe_B}
\\
&&\frac{\chi}{r^2}\left(-1+B +rB\frac{N'}{N}\right)
=8\pi Gp-\chi^2V
\nonumber\\&&\qquad\qquad\qquad\qquad\qquad
-B\left(\frac{2}{r}+\frac{N'}{2N}\right)\chi'.
\label{fe_N}
\end{eqnarray}
The equation of motion for $\chi$ [Eq.~(\ref{scalareq})] gives
\begin{eqnarray}
&&B\left[\chi''+\left(\frac{2}{r}+\frac{N'}{2N}+\frac{B'}{2B}\right)\chi'\right]
\nonumber\\&&\qquad\qquad\qquad
=\frac{8\pi G}{3}(-\rho+3p)+\frac{2\chi^3}{3}\frac{dV}{d\chi}.
\label{eq_chi}
\end{eqnarray}
We do not integrate the angular components of the field equations.
Instead, we use them
to check the accuracy of our numerical results, because
those are derived from other equations via the Bianchi identity.

If the energy density is constant inside the star, $\rho=\rho_0$,
Eq.~(\ref{dp=}) immediately gives
\begin{eqnarray}
N(r) = \left[\frac{\rho_0+p_c}{\rho_0+p(r)}\right]^2.
\end{eqnarray}
In the main text we only consider constant density stars for simplicity.

\subsection{Boundary conditions}

Let us study
the boundary conditions at the center of a star.
Assuming the regularity,
we expand the variables in the power series of $r$ as
\begin{eqnarray}
&&N(r)=  1+ N_2 r^2+... ,\quad
B(r)= 1+ B_2 r^2+...,\quad
\nonumber\\
&&
\chi(r)= \chi_c\left(1+ \frac{C_2}{2}r^2+...\right),\label{series-r=0}
\\
&& 
\rho(r)=\rho_c +{\rho_2\over 2}r^2+ ... ,\quad
p(r)= p_c+\frac{p_2}{2} r^2+... ,
\nonumber
\end{eqnarray}
where $\chi_c$,  $\rho_c$ and $p_c$ 
are the central values of the scalar field, the energy density and 
the pressure, respectively.
Note that using the scaling freedom of the $t$ coordinate, we set
 $N(0)=1$.
From Eqs.~(\ref{fe_B})--(\ref{eq_chi}), we obtain
\begin{eqnarray}
3 B_2&=&-8\pi \hat{G}\rho_c -\chi_c V_c -3C_2,
\\
B_2+2N_2& =& 8\pi \hat{G}p_c-\chi_c V_c-2C_2,
\\
3C_2& =& \frac{8\pi \hat{G}}{3}(-\rho_c+3p_c)+\frac{2\chi_c^2}{3}V_{\chi_c},
\end{eqnarray}
where $\hat{G}:=G/\chi_c$,
$V_c:=V(\chi_c)$, and $V_{\chi_c}=dV/d\chi|_{\chi=\chi_c}$.
These three equations are rearranged to give
\begin{eqnarray}
B_2&=&-\frac{8\pi \hat{G}}{9}\left(2\rho_c+3 p_c\right)
-\frac{\chi_c}{3}V_c-\frac{2\chi_c^2}{9}V_{\chi_c},\label{ex-B2}
\end{eqnarray}
\begin{eqnarray}
N_2&=&\frac{8\pi \hat{G}}{9}(2\rho_c+3p_c)-\frac{\chi_c}{3}V_c
-\frac{\chi^2_c}{9}V_{\chi_c},
\\
C_2&=&\frac{8\pi\hat{ G}}{9}(-\rho_c+3p_c)+\frac{2\chi_c^2}{9}V_{\chi_c}.
\end{eqnarray}
Then, $p_2$ is derived from the conservation equation:
\begin{eqnarray}
p_2+N_2\left(\rho_c+p_c\right)=0.\label{ex-p2}
\end{eqnarray}
The Ricci scalar is given by
$R= R_c+{\cal O}(r^2)$ with $R_c=-6(B_2+N_2)$ near $r=0$.


\end{document}